\documentclass[aps,prl,twocolumn,amsmath,amssymb,floatfix,showpacs,superscriptaddress]{revtex4}
\usepackage{amsmath,amssymb,amsfonts,mathrsfs}
\usepackage{color}
\usepackage{epsfig}
\usepackage{psfrag}
\usepackage{graphicx}
\usepackage{color}

\begin{document}

\title{Charge Fractionalization in a Mesoscopic Ring}

\author{Wade DeGottardi$^1$, Siddhartha Lal$^1 \ ^2$, and Smitha Vishveshwara$^1$}
\affiliation{\small{
$^1$Department of Physics, University of Illinois at
Urbana-Champaign, 1110 W.\ Green St.\ , Urbana, IL  61801-3080, USA\\
$^2$Department of Physical Sciences, IISER-Kolkata, Mohanpur Campus, West Bengal - 741252, India
}}

\date{\today}

\begin{abstract}{We study the fractionalization of an electron tunneling into a strongly interacting electronic one-dimensional ring. As a complement to transport measurements in quantum wires connected to leads, we propose non-invasive measures involving the magnetic field profile around the ring to probe this fractionalization.
In particular, we show that the magnetic-field squared produced by the electron and
the power that it would induce in a detector exhibit anisotropic profiles that depend on the
degree of fractionalization. We contrast true fractionalization with two other scenarios which
could mimic it -- quantum superposition and classical probabilistic electron insertion. We
show that the proposed field-dependent measures and those of the persistent current can distinguish between these three scenarios.}
\end{abstract}

\maketitle

\begin{figure}[htb]
\begin{center}
\includegraphics[bb=100 0 400 300,scale=0.4]{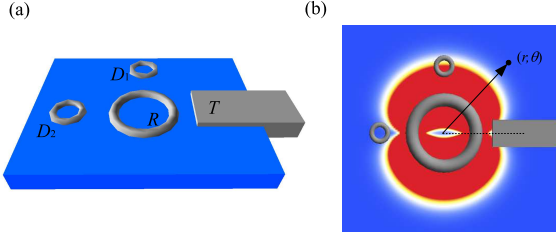}
\end{center}
\caption{(a) Oblique view  of the setup in which a clockwise-moving electron injected into the ring from tunnel junction $T$ fractionalizes into clockwise- and counterclockwise-moving quasiparticles. (b) Top-down view  of the same setup with an overlay of a spatial plot of $\overline{\langle B^2 \rangle}$ predicted for a ring with Luttinger parameter $g = 0.2$.}
\label{fig:setup}
\end{figure}

A spectacular feature of strongly correlated
low-dimensional electronic systems is that collective behavior renders the electron completely unstable, resulting in its fractionalization~\cite{review}. As a prime example, in a one-dimensional quantum wire, Tomonaga-Luttinger liquid (TLL) theory predicts that a momentum-resolved electron tunneling into the wire splinters into charges $(1\pm g)e/2$ moving in opposite directions, where $g$, the Luttinger
parameter depends on the ratio of interactions and Fermi energy and is unity in the absence of interactions~\cite{safi,pham}. Exciting developments in experimental capabilities have enabled the physical realization of such a situation~\cite{steinberg}. These studies inspire revisiting fractionalization in a new light and addressing a spectrum of theoretical and physical issues. For instance, can one distinguish true fractionalization from quantum mechanical probabilistic processes? Or even classical probabilities? Are there geometries which could eliminate one of the biggest banes in detecting fractionalization --
the effect of leads~\cite{safi,stone,pugnetti,kim,frequency}? What measurements in such
geometries could pinpoint true fractionalization?  In this Letter, we
answer each of these questions in the context of the ring geometry
illustrated in Fig.~\ref{fig:setup}.

Here, an electron tunnels from a lead into a thin mesoscopic ring and, as with the quantum wire~\cite{leurtheory}, has a well-defined momentum profile. Strong interactions within the ring cause an electron associated, for instance, with clockwise Fermi-momentum to decompose into two components of charge $(1\pm g)e/2$ moving in clockwise (CW) and counter-clockwise (CCW) directions. Our study focuses on the magnetic field produced by such a situation and the signatures of fractionalization reflected in the spatial distribution of higher moments involving this field. Specifically, we propose measurements of the field squared, as for instance can be measured by a SQUID, and the power induced by the field in a pickup loop (see Fig.~\ref{fig:setup}). These measurements have the advantage of purely entailing d.c. quantities as opposed to high frequency measurements and of constituting weak, i.e., non-invasive, read-outs when compared with those involving lead attachment.

Fractionalization emerges from the strongly correlated nature of the many-body wavefunction and is fundamentally different from quantum mechanical superpositions or classical probabilities involving individual particles even though these processes can mimic one another in measurements. In our situation, we explore three different scenarios that represent each of these situations. First, the fractionalized state resulting from the tunneling of a CW moving electron $\psi^\dagger_+(x)$ having a wave function spread  $\chi(x)$ above the TLL ground state, $|G \rangle_{LL}$, is given by $| F \rangle = \int \chi(x) \psi^\dagger(x) dx |G \rangle_{LL}$.  Second, a quantum superposition state, $| QS \rangle$, that mimics the fractionalized state would consist of superpositions of CW$(+)$/CCW$(-)$ {\it electrons} excited above a non-interacting Fermi gas ground state $|G\rangle_0$, i.e. $| QS \rangle = \sum_{\pm} f_\pm\int \chi(x) \psi_\pm^\dagger(x) dx |G \rangle_{0}$, where $f_\pm=\sqrt{(1\pm g)/2}$ correspond to the mimicking probabilities. Third, a classical probabilistic situation would correspond to an ensemble of CW and CCW electrons excited in the non-interacting Fermi gas, denoted by the density matrix $\mathbf{M}_\rho = \sum_\pm f^2_\pm | \pm \rangle \langle \pm |$, where $| \pm \rangle = \int \chi(x) \psi_\pm^\dagger(x) dx | G \rangle_0$. In what follows, after introducing fractionalization in the TLL liquid ring setting, we show that a combination of the two magnetic field measures combined with persistent current signatures in the mesoscopic ring, at once distinguish the three different scenarios and provide a means of extracting the Luttinger parameter.

To briefly summarize TLL physics in a ring geometry (see for example,~\cite{loss}),
we consider a one-dimensional system with position $x$ denoting the
circumferential direction bounded by  $0 \leq x < 2 \pi R$, where $R$ is the radius of the
ring. The ring geometry imposes periodic boundary conditions on
electron operators  such that $\psi(x+2 \pi R) = e^{i 2 \pi\Phi/\Phi_0}
\psi(x)$ where $\Phi$ is any flux threading the ring and $\Phi_0 =
h/e$.
For electrons filling a Fermi sea, we decompose the electron operators as
 $\psi(x) = \sum_r \psi_r(x)$ where
$\psi_r^\dagger$ denotes the creation operator for a $r=\pm$ moving electron.
The kinetic energy for linearized
low-energy modes moving at a Fermi velocity $v_F$
takes the form $H_0 = - i v_F \int dx \sum_r \psi_r^\dagger
\partial_x^{\phantom\dagger} \psi_r$. As is commonly done, we
restrict interaction effects to the short-range form $H_{int} = V
\int dx \ \rho^2(x)$ where $\rho = e \sum_{r=\pm} \rho_r$ is the sum of
charge densities $\rho_r = \psi_r^\dagger \psi_r^{\phantom\dagger}$.
Of physical interest,  the current operator is given by $\hat{I} = e v_F j$, where $j = \sum_r r \rho_r$.

This model is amenable to a bosonization treatment via the
transformation $\psi_r(x) \sim e^{i r k_F x} e^{i \sqrt{\pi}
\varphi_r(x)}$ giving $\rho_r = k_F/2 \pi + r \partial_x \varphi_r / \sqrt{2 \pi}$, where the chiral bosonic fields $\varphi_r$ satisfy
the commutation relations $[\varphi_r(x), \varphi_{r'}(x')] =  i
r \delta_{r r'} \mbox{sgn}(x-x')$ and $k_F$ is the Fermi momentum.
The net Hamiltonian $H_0 + H_{int}$ may be brought into the free TLL form
via a Bogoliubov transformation of the $\phi$ fields~\cite{bosonize}, yielding
\begin{equation}
 H_{LL} = \frac{u}{4} \int dx \left[ \left(\partial_x \tilde{\varphi}_+ \right)^2 +  \left(\partial_x \tilde{\varphi}_- \right)^2 \right],
\label{eq:HamLL}
\end{equation}
 where $u=v_F/g$ is the the plasmon
velocity, $g$ is the Luttinger parameter $g \equiv 1/\sqrt{1 + 2 V/\pi \hbar v_F}$ and $\tilde{\varphi}_\pm(x \mp
ut)$ are transformed chiral bosons.

The fractionalization of an electron can be seen by representing an electron operator having CW Fermi momentum in terms of the chiral bosons:
\begin{equation}
\psi_+^\dagger(x,t) \sim e^{- i k_F x} e^{-i \frac{\sqrt{\pi}}{2 \sqrt{g}} \left[ \left( 1 + g \right) \tilde{\varphi}_+(x,t) + \left( 1 - g \right) \tilde{\varphi}_-(x,t) \right]}.
\label{eq:elecfrac}
\end{equation}
By relating the chiral bosonic fields to the charge and current density operators, $\rho$ and $j$, respectively, it follows that the operator $e^{-i\sqrt{\frac{\pi}{g}}\tilde\varphi_+(x,0)}$ creates a unit charge $e$ that at time $t$ can be found at position $x-ut$. Thus, we see that the electron operator in Eq.~\ref{eq:elecfrac} creates the fractional charges $(1\pm g)e/2$ moving in opposite directions. More explicitly, in the situation of interest, the state of the ring after the injection of a CW moving electron at time $t=0$ is given by $| F
\rangle =  \int \chi (x) \psi_{+}^\dagger(x,t=0) dx|G \rangle_{LL}$ ~\cite{leinaas}.
It is straightforward to calculate the expectation value of the
current in this state, $I(x,t) = \langle \hat{I}
\rangle_F \equiv \langle F | \hat{I}(x,t) | F \rangle$, yielding
\begin{equation}
I(x,t) = \frac{e u}{4\pi R} \left[ (1+g) | \chi(x - u t) |^2 + (1-g) |\chi(x+ut)|^2 \right].
\label{eq:currentexp}
\end{equation}
The form of the current explicitly demonstrates that the electron  splinters  into two
components that rotate in opposite directions, have the same profile $\chi(x)$ and carry
charges $(1\pm g)e/2$.

The magnetic field produced by these counter-propagating charges can
be evaluated by using the Biot-Savart law to define the magnetic
field operator at position ${\bf r}$ as $\hat{\mathbf{B}} =
\frac{\mu_0}{4 \pi} \int d\ell \ \hat{ \mathbf{I}}(\ell) \times
\mathbf{r}/ | \mathbf{r} |^3 $, where $\mu_0$ is permeability of free space. At any
given point having polar coordinates $(r,\theta)$ in the plane of the
ring, where the origin is at the ring's center and the electron is
inserted at $(R,0)$, the current in Eq.~\ref{eq:currentexp} produces
a field perpendicular to the plane. For the case of $\chi$
having a spread much smaller than the ring diameter, the
$z$-component of the field takes the form
\begin{equation}
\langle \hat{B}_z \rangle_F = \frac{\mu_0 e\omega}{2R} \left[ (1+g) h(t) - (1-g) h(-t) \right],
\label{eq:Bz}
\end{equation}
where $\omega = u/R$ and $h(t)=(1-a(t))/(\frac{r^2}{R^2}-2a(t)+1)^{3/2},
a(t)=r\cos(\omega t -\theta)/R$. In principle, a time-resolved measurement
of the magnetic field, as with other quantities, such as conductance, would yield information on fractionalization.
However, as is the goal here, we seek low-frequency or
time averaged signatures. Although the tunneling of the electron picks out a
specific point on the ring, signatures are effaced by time-averaging
of any quantity that is linear in the ring current. For example,
$\overline{\langle B_z \rangle}$ shows an isotropic spatial profile.
(Here, we use the overline to denote time average.)

We thus focus on two measures that are quadratic in the current and can be
obtained from a continuous weak linear measurement~\cite{CWLM} via inductive
coupling to the ring. The first is simply $S(r,\theta) = \overline{\langle B^2_z \rangle}$,
which can be accessed in a SQUID detector biased to a minimum of its
I-V characteristic curve. The second is the  average power received
by a detector, for example, an ultra-sensitive bolometer. For a small conducting
detector (ignoring local spatial variations in the magnetic field),
this is given by $P(r,\theta) = \overline{ \langle\partial_t B_z
\rangle^2}$. Crucially, note that the former involves a quantum
average of a quadratic operator and the latter that of a linear
operator.

\begin{figure}[htb]
\begin{center}
\includegraphics[bb=0 0 640 480,scale=0.5]{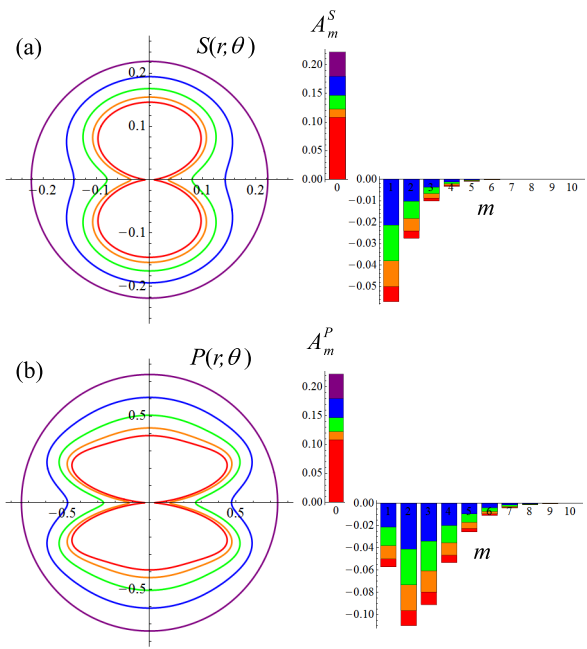}
\end{center}
\caption{Polar plots of (a) $S(r,\theta) = \overline{\langle B_z^2(t) \rangle}$ and (b) $P(r,\theta) = \overline{\langle \partial_t B_z(t) \rangle^2}$
at $r = 2R$ for values of the Luttinger parameter $g =  1.0,\ 0.8,\ 0.6,\ 0.4,\ 0.2$ (from outermost and most isotropic to the innermost) as a function of $\theta$.  Bar graphs of the spectral weight of the maps (c) $S$ and (d) $P$ showing the even Fourier coefficients as defined in Eq.~\ref{eq:FourierSP}, i.e., the height of columns for $m = 0, 1, 2$ correspond to the zeroth, $\cos 2 \theta$, and $\cos 4 \theta$ terms, respectively (the zeroth  and non-zeroth coefficients are
shown on a different scale). With increasing fractionalization (decreasing $g$), spectral weight is transferred to the non-zeroth coefficients reflecting increasing anisotropy.}
\label{fig:templates}
\end{figure}

The forms of the moments $S$ and $P$ can be easily evaluated by
taking appropriate quantum expectations ($\langle \rangle$) and time averages (overline) to obtain $S(r,\theta) = \left( \frac{\mu_0 e \omega}{2R}
\right)^2 \left[ (1+g^2) \overline{h^2(t)} - (1-g^2) \overline{ h(t)
h(-t)} \right]$ and a similar form for $\tilde{P}\equiv P/\omega^2$ with $h(t)$ replaced by its time derivative $h'(t)$. Information on fractionalization is best analyzed by resolving these quantites into their angular Fourier coefficients:
\begin{equation}
S/\tilde{P}(r,\theta) = \left( \frac{ \mu_0 e \omega}{2R} \right)^2 \sum_{m = 0}^\infty A^{S/P}_m(r) \cos 2 m \theta.
\label{eq:FourierSP}
\end{equation}
In the non-interacting ($g=1$) limit, an electron circles the ring in the CW direction, retaining rotational symmetry on average and thus we have
$A^{S/P}_m=0, m \neq 0$.

The plots in Fig.~\ref{fig:templates} capture our central
result that higher moments of the current and of the magnetic field
profile (in our case, $S$ and $P$) reflect the concurrent motion of
fractionalized charge components in their rotational symmetry
broken distributions. In the explicit forms of $S$ and $P$ above, given that
$\overline{h^2(t)}$ preserves rotational symmetry while $\overline{ h(t)
h(-t)}$ breaks it, we see that primarily $A^{S/P}_0$ scale as $1+g^2$ and
$A^{S/P}_{m\neq 0}$ as $1-g^2$. This distribution is illustrated in the plots
of Fig. ~\ref{fig:templates} and also agrees with the rotationally symmetric
non-interacting limit ($g=1$). The bilateral symmetry of the plots reflects
the two charge components moving away from the injection point and towards
the diametrically opposite point. That these two special points exist for
any arbitrary closed shape suggests that our result that fractionalization
causes a distribution that distinguishes two points is robust for any closed
loop.

We contrast the behavior of the moments $S$ and $P$ in the
fractionalized state to the quantum and classical probabilistic
situations. In the quantum state $| QS \rangle$,  a superposition
of CW and CCW moving electrons, quantum averages
of operators that are linear in the current mimic charge fractionalization
while their higher moments, (for example, $\langle \hat{B}^2_z\rangle $)those of
linear operators do not. Thus, $S = \overline{\langle \hat{B}^2_z \rangle}$ is
isotropic but $P = \overline{ \langle\partial_t \hat{B}_z
\rangle^2}$ shows an anisotropic profile similar to that of Fig.
~\ref{fig:templates}. b. For the classical situation described by
the density matrix $\mathbf{M}_\rho$, the moments are evaluated by
separately considering CW and CCW electrons and adding their
appropriately weighted contributions. Thus, both moments yield
isotropic profiles. As summarized in Table. I, the two measurements
therefore can distinguish between the three possible scenarios.

In addition to the differences mentioned here, Ref.~\cite{leinaas}
distinguishes true fractionalization from other situations by the profound observation that charge fluctuations are in fact a feature of the many-body groundstate and the background of particle-hole excitations while the fractionalized electron is itself `sharp'. Translated to our setting, we expect fluctuations in the magnetic field to be induced even by the quiescent TLL ring (having no extra tunneled electron) and identical to those induced by the fractionalized state $|F\rangle$.

\begin{figure}[htb]
\begin{center}
\includegraphics[bb=100 30 400 270,scale=0.4]{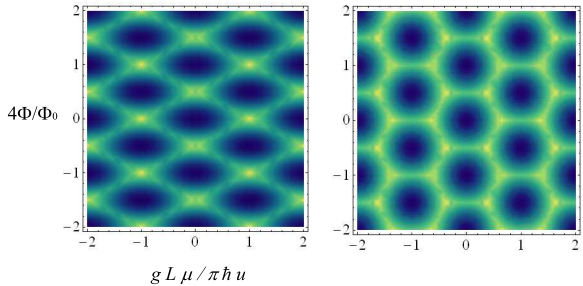}
\end{center}
     \caption{Ground state structure obtained from the Hamiltonian $H_{NJ}$ of Eq.~\ref{eq:CB}. Each cell corresponds to a given electron number ($N$) and persistent current ($J$) which optimizes $H_{NJ}$ as a function of chemical potential (horizontal axis) and magnetic flux (vertical); bright lines indicate a transition in which $J$ or  $N$ changes by 1. Cells are  (a) diamond shaped for $g = 1$ (non-interacting system) and (b) hexagonal shaped for  $g (= 1/2) \neq 1$ with horizontal side length $1-g^2$.}
     \label{fig:CB}
\end{figure}

Thus far, we have described the injection localized electron wavepacket as a superposition of plasmon-like modes described by Eq.~\ref{eq:HamLL}. Typically, due to coupling to the environment~\cite{CedraschiButtiker,open,safisaleur},
these modes have a finite lifetime, giving rise to a characteristic decoherence time $\tau_d$
within which measurements need to be performed.
However, for timescales longer than $\tau_d$, no plasmons are excited and tunneled electrons need purely be described by the excess electron number, $N = \sum_r n_r$, and the persistent current due to the number imbalance between CCW- and CW-moving electrons, $J = \sum_r r n_r$, where we define $n_r = r \Phi/\Phi_0 + \int dx \ \rho_r$. The optimal values of these `topological' quantities $N$ and $J$ can be tuned by the application of a gate potential $\mu$ and external flux
$\Phi$, and can be determined by minimizing the energy functional derived from Eq.~\ref{eq:HamLL}\cite{kinaret}
\begin{eqnarray}
H_{NJ} = \frac{\pi \hbar u}{4\pi R} \left[
\frac{1}{g} N^2 + g \left( J + \frac{2\Phi}{ \Phi_0} \right)^2
\right] - \mu N. \label{eq:CB}
\end{eqnarray}
The regions of different optimal $N$ and $J$ values can be charted by Coulomb blockade measurements wherein conductance peaks track electron occupation numbers on the ring. We show the boundaries for these regions in (dimensionless) $\mu$-$\Phi$ parameter space in Fig.\ref{fig:CB}. Interactions render these regions to be generically hexagonal, characterized by horizontal sides of length $1-g^2$. Thus, the geometry of this diagram is an easily accessible, alternate means of extracting $g$, the Luttinger parameter.

 A highlight of this slow-time regime is that it offers another route to distinguishing the fractionalized state $| F \rangle$ by way of persistent current analysis. Ultimately this state is associated with a CW electron and hence has the fixed current value $J=+1$ while the quantum and classical states characterized by $| QS \rangle$ and $\mathbf{M}_\rho$ involve CW and CCW electrons, thus showing values $J=\pm 1$ which vary between measurements. Thus, as summarized in Table I, the anisotropy in moment $S$ and non-variability in persistent current distinguish the fractionalized state
from the quantum and classical scenarios (though the latter is not a smoking gun test) while anisotropy in the moment $P$ distinguishes classical scenario.

Finally, to  provide relevant estimates for experiments,
for radius $R \approx 1 \ \mu$m and typical circulating frequency $\omega \approx 10^{11}$ Hz, we have $\left(\mu_0 e
\omega /2R\right)^2 \approx \left(0.1 \mbox{ milligauss}\right)^2$
and $\left(\mu_0 e \omega^2 /2R\right)^2 \approx \left(80 \mbox{
T/sec}\right)^2$. An important requirement is that the
injection of an electron must be made on a timescale $\tau_T \ll
1/\omega$ in order for there to be a `clean' injection of the
electron. For the ring,
we have $\tau_T = R_T C$ where $R_T$ is the tunnel junction
resistance and $C$ ($\sim \epsilon_0 R \sim 10^{-17}$ F) is the ring capacitance.
This gives the requirement that $R_T \ll 1 \ $M$\Omega$. On the other
hand, the Coulomb blockade limit holds only if $R_T \gg \frac{h}{e^2}
= 26$ k$\Omega$. Thus, we need a $R_T \sim 100$ k$\Omega$. Another consideration
is that interaction effects at the tunneling point restrict the energy window
in which our results hold~\cite{safisaleur}. The role of the electron's spin can also come
into play and can be analyzed by a simple generalization of our results.

In conclusion, we have presented an alternative to the quantum wire based
electron-in electron-out paradigm for charge fractionalization in the arena of weak measurements in
mesoscopic rings. Our envisioned setup discerns subtle attributes that distinguish fractionalization from quantum and classical probabilistic scenarios and is within the reach of current nanotechnology.

\begin{table}
\caption{Results of various measurements exemplifying how fractionalization ($|F\rangle$) in a TLL can be differentiated from the quantum ($|QS\rangle$) and classical scenarios ($\mathbf{M}_\rho$) considered in the text. Time-averaged quantities $S$ and $P$ can display isotropic (I) or anisotropic (AI) distributions. Persistent current measurements can yield non-variable (NV) or variable (V) outcomes for repeated measurements.}
\begin{center}
\begin{tabular}{|r|r|r|r|}
  \hline
  & $| F \rangle $ &  $| QS \rangle$ & $\mathbf{M}_\rho$ \\
  \hline
  \hline
  $S(r,\theta) = \langle \hat{B}^{2}_{z} \rangle $ & AI & I & I  \\
  \hline
  $P(r,\theta) = \langle \partial_t \hat{B}_z\rangle^2$ & AI  & AI & I   \\
  \hline
  Persistent Current & NV &  V  & V    \\
  \hline
\end{tabular}
\end{center}
\end{table}

\section{Acknowledgments}

We are grateful to  Raffi Budakian and Charles Kane for their perceptive comments. For their support, we thank the NSF under grant DMR 0644022-CAR (W.D. and S.V.), the CAS fellowship at UIUC (S.V.) and the DST, Govt. of India, under a Ramanujan Fellowship (S.L.).


\begin{thebibliography}{99}

\bibitem{review} V. V. Deshpande, M. Bockrath, L. I. Glazman, A. Yacoby, Nature {\bf 464} 209-216 (2010).

\bibitem{safi} Safi, I and Schulz, H. J. Phys. Rev. B {\bf 52}, R14265 (1996).


\bibitem{pham} K.-V. Pham, M. Gabay, and P. Lederer, Phys. Rev B
    {\bf 61}, 16397 (2000).


\bibitem{steinberg} H. Steinberg et al. Nature Physics {\bf 4},
    116 (2008).


\bibitem{stone} D. L. Maslov and M. Stone, Phys. Rev. B {\bf 52}, R5539 (1995).

\bibitem{frequency} V. V. Ponomarenko, Phys. Rev. B {\bf 54}, 10328 (1996).


\bibitem{pugnetti} S. Pugnetti et al. Phys. Rev. B, {\bf 79}, 035121
    (2009).

\bibitem{kim} J. U. Kim, W.-R. Lee, H.-W. Lee, H.-S. Sim, PRL {\bf 102}, 076401 (2009).

\bibitem{leurtheory} K. Le Hur, B. I. Halperin, A. Yacoby, Annals of Physics {\bf 323}, 3037 (2008).


\bibitem{loss} Loss, D. PRL {\bf 69}, 343 (1992).


\bibitem{bosonize} M. Stone, \emph{Bosonization}. World Scientific, Singapore (1994).

\bibitem{CWLM} D.V. Averin, Physica C {\bf 352}, 120, (2001).

\bibitem{leinaas} J. M. Leinaas, M. Horsdal, T. H. Hansson, Phys. Rev. B {\bf 80}, 115327 (2009).

\bibitem{safisaleur} I. Safi and H. Saleur, PRL {\bf 93}, 126602 (2004).

\bibitem{open} A. H. Castro Neto, C. de C. Chamon, C. Nayak, Phys. Rev. Lett. {\bf 79}, 4629 (1997).

\bibitem{CedraschiButtiker} P. Cedraschi, V. V. Ponomarenko and M. Buttiker, PRL {\bf 84}, 346 (2000).

\bibitem{kinaret} Kinaret, Jonson, Shekhter, and Eggert. Phys. Rev. B {\bf 57}, 3777 (1998).



\end{thebibliography}
\end{document}